# Recovering Residual Forensic Data from Smartphone Interactions with Cloud Storage Providers

**George Grispos**
University of Glasgow
g.grispos.1@research.gla.ac.uk

**William Bradley Glisson**
University of South Alabama
bglisson@southalabama.edu

**Tim Storer**
University of Glasgow
Timothy.Storer@glasgow.ac.uk

## Abstract[1]

There is a growing demand for cloud storage services such as Dropbox, Box, Syncplicity and SugarSync. These public cloud storage services can store gigabytes of corporate and personal data in remote data centres around the world, which can then be synchronized to multiple devices. This creates an environment which is potentially conducive to security incidents, data breaches and other malicious activities. The forensic investigation of public cloud environments presents a number of new challenges for the digital forensics community. However, it is anticipated that end-devices such as smartphones, will retain data from these cloud storage services. This research investigates how forensic tools that are currently available to practitioners can be used to provide a practical solution for the problems related to investigating cloud storage environments. The research contribution is threefold. First, the findings from this research support the idea that end-devices which have been used to access cloud storage services can be used to provide a partial view of the evidence stored in the cloud service. Second, the research provides a comparison of the number of files which can be recovered from different versions of cloud storage applications. In doing so, it also supports the idea that amalgamating the files recovered from more than one device can result in the recovery of a more complete dataset. Third, the chapter contributes to the documentation and evidentiary discussion of the artefacts created from specific cloud storage applications and different versions of these applications on iOS and Android smartphones.

## 1. Introduction

An increase in demand for information technology (IT) resources has prompted many organizations to turn their attention to cloud computing. This technology has significant potential to reduce costs and increase efficiency in the workplace [1]. Migrating to a cloud computing environment means an organization can replace much of its traditional IT hardware with virtualized, remote and on-demand infrastructure services such as storage space, processing power and network bandwidth [2].

Storing corporate data online using cloud-based storage services such as Amazon S3, Google Docs and Dropbox has become an effective solution for the business needs of a growing number of organizations [3]. Cloud storage services can offer an organization greater flexibility and availability, with virtually unlimited storage space, as well as the ability to synchronize data between multiple devices. Typically, cloud storage providers will operate on the 'freemium' financial model, offering customers free storage space with an option to purchase further unlimited storage space as they require [4, 5]. This business model has demonstrated to be successful, as the popularity of cloud storage services has soared in recent years. For example, Dropbox has seen its customer-base surpass 300 million users and now claim that over 1 billion files are saved every three days using its services [6, 7]. Mozy claim that more than six million individual users and 100,000 businesses are using their services, while Box has reported that implementation of its mobile device application increased 140% monthly in 2011 [10, 11]. Forrester Research has predicted that approximately two-thirds of adults who use the Internet in the United States are using some form of personal cloud storage service, often combining cloud services for both work and personal use [12]. Cloud storage services are also increasingly integrating into the retail, financial, legal and healthcare enterprise markets [13, 14].

---
[1] **Please cite this paper as:** George Grispos, William Bradley Glisson and Tim Storer. (2015) Recovering residual forensic data from smartphone interactions with cloud storage providers, In The Cloud Security Ecosystem, edited by Ryan Ko and Kim-Kwang Raymond Choo, Syngress, Boston, Pages 347-382.

Although the benefits of using cloud storage services are attractive, a major concern is the security and privacy of data in these environments [15-18]. For many organizations, there are several reasons to decline adoption of cloud storage services, including a necessity to protect mission-critical information, legal and regulatory obligations and concerns regarding the confidentiality and integrity of their information [19-23]. These reservations are being validated as TrendMicro has reported that cloud adopters have witnessed an increase in the number of cloud security incidents as compared to traditional IT infrastructure security events [24]. Further complicating matters, researchers have demonstrated how cloud storage services, such as Dropbox, can be hijacked and exploited to gain access to an unsuspecting user's account [4, 25]. There is no practical barrier preventing further exploitation of cloud storage services by users to access and retrieve files. Security incidents and criminal activity involving cloud storage services could require a subsequent forensic investigation to be undertaken.

There is a general consensus from both industry and academia that it may be difficult to investigate inappropriate or illegal activity involving cloud computing environments [2, 26-29]. One of the biggest challenges for forensic investigators examining cloud-based services is the ability to identify and recover digital evidence in a forensically sound manner [2, 28]. This problem is magnified in public cloud environments, such as those used by cloud storage providers. The remote, distributed and virtualized nature of a public cloud environment means that the conventional, offline approach to forensic evidence acquisition is largely invalidated [2]. The tools and methods used to preserve and acquire a forensic copy of data stored on a traditional storage device are unlikely to transfer to a public cloud environment [2]. The remote and distributed nature of public cloud architectures can also make the identification of a single storage device containing relevant data impractical and even impossible. This means that an investigator cannot directly obtain a copy of the evidence required for analysis. An alternative approach is for the investigator to request the cloud storage provider to obtain a forensic copy of the storage device. However, this approach can take a significant amount of time, or be obstructed by cross-border jurisdictional disputes [22, 30].

This research investigates a practical solution to problems related to investigating cloud storage environments using practitioner-accepted forensic tools. This study extends the results from an initial investigation, which examined the feasibility of an end-device, providing a proxy view of the evidence in a cloud forensics investigation [31]. Relevant information and data from that conference publication has been included in this chapter for completeness. The contribution of this chapter is threefold. First, the findings from this research further supports the idea that end-devices which have been used to access cloud storage services can be used to provide a partial or snap-shot view of the evidence stored in the cloud service. Second, the chapter provides a comparison of the files which can be recovered from different versions of cloud storage applications. The chapter also supports the idea that amalgamating the files recovered from more than one device can result in an investigator recovering a more complete 'dataset' of files stored in the cloud service. Third, the chapter contributes to the documentation and evidentiary discussion of the artefacts created from specific cloud storage applications and different versions of these applications on iOS and Android smartphones.

The chapter is structured as follows: Section 2 discusses the challenges of conducting digital forensic investigations in a cloud computing environment and examines what work has been done in relation to cloud storage forensics, as well as presenting an overview of smartphone forensics. Section 3 proposes the hypotheses and research questions which guided this research and describes the experimental design undertaken to address the research questions. Section 4 reports the findings, and Section 5 is used to discuss the results and their impact on forensic investigations. Finally, Section 6 draws conclusions from the work conducted and presents future work.

## 2. Related Work

A growing number of researchers have argued that cloud computing environments are inherently more difficult to investigate than conventional environments [2, 26-28, 32]. Ruan, et al., [27] defined the term 'cloud forensics' as a "cross discipline of cloud computing and digital forensics" and described cloud forensics as a subset of network forensics. However, this definition does not take into consideration the virtualization aspect of the cloud [2]. Ruan, et al., [27] also noted that an investigation involving cloud computing would include technical, organizational and legal aspects. Grispos, et al., [2] described how traditional digital forensic models and techniques used for investigating computer systems could prove to be ineffective in a cloud computing environment. Furthermore, Grispos, et al., [2] identified several challenges for forensic investigators including: creating adequate forensic images, the recovery of segregated evidence and large data storage management. Taylor, et al [28] raised the concern that potential important evidence could be lost in a cloud environment. Registry entries in Microsoft Windows platforms, temporary files, and metadata could all be lost if the user leaves the cloud [28]. Reilly, et al., [26] speculated that one potential benefit of investigating a public cloud environment is that the data being investigated will be located in a central location, which means that incidents can, potentially, be investigated quicker. This is unlikely to be the case as the very nature of a public cloud service theoretically means that even evidence related to individuals within the same organization could be segregated in different physical locations and stored alongside data belonging to other organizations and the general public [33].

To enable a forensic investigation to be conducted, evidence needs to be collected from cloud computing environments, thus introducing a unique set of challenges for forensic investigators [34]. Researchers have begun proposing methods of acquiring evidence from a variety of cloud providers and services [34, 35]. Delport, et al., [35] proposed the idea of isolating a cloud instance for further investigation, however, it is not clear how a forensic image of the instance under investigation is obtained after it has been isolated from the rest of the cloud environment. Dykstra and Sherman [34] proposed three methods of evidence collection from Infrastructure-as-a-Service (IaaS) instances stored in the Amazon EC2 Cloud. The first method evaluated the performance of several forensic tools including FTK Imager and Encase Enterprise, which were used to extract evidence directly from the cloud instance in the Amazon Cloud. An issue with this method is that the investigator must be in possession of Amazon EC2 key pairs used to connect to the instance. The purpose of the key pairs is to ensure that only the instance's owner has access to the instance [36]. These public/private keys are created by the owner when the instance is first created using the Amazon Web Services Management Console [37]. Unless the investigator can recover these keys, this method of acquisition cannot be used. The second proposed method involved acquiring evidence from the virtualization layer of a cloud by injecting a remote agent into the hypervisor of the cloud environment. This approach was evaluated in a private cloud environment, where the investigator had the ability to write into memory the guest virtual instance. However, this is unlikely to be the case in a public cloud environment, where the cloud provider will control access to the hypervisor and virtualization layer of a cloud [19]. The final method proposed by Dykstra and Sherman involves requesting Amazon EC2 to collect the required evidence from the host on behalf of the investigator. Dykstra and Sherman note that the limitation of this method is that Amazon does not provide checksums to verify the integrity of the forensic image; therefore, the investigator cannot be certain that the data supplied by Amazon and the data stored in the cloud are identical [34].

Researchers have also attempted to define investigative frameworks specifically addressing cloud storage providers [38-42]. Lee, et al. [39] presented a framework for investigating incidents involving the Apple iCloud environment. The framework proposes the idea that the forensic investigator can examine Windows and Apple OSX-based systems, as well as Apple mobile devices to recover traces of data stored in the iCloud service. The research focused on the recovery of email messages, memos, contacts, calendar information and bookmarks. The limitation of this research is that Lee, et al. [39] did not examine or discuss if the artefacts recovered from the end-devices were representative of the data stored in the cloud. There is the possibility that evidence still stored in the iCloud service was not recovered from the end-devices. Chung, et al. [38] have also proposed a framework for investigating cloud storage services including Amazon S3, Google Docs, Dropbox and Evernote.

Quick and Choo [40-42] have undertaken three separate case studies to investigate the data remnants from Dropbox, Google Drive and SkyDrive (now called OneDrive). In all three case studies, a Windows 7 personal computer and an Apple 3G iPhone were used to access and view a dataset stored in the cloud storage service. A Windows 7 personal computer was emulated using virtual machines. A variety of web browsers were used in conjunction with the virtual machine and the specified cloud storage client application to collect data [40-42]. The findings from the personal computer analysis revealed that usernames, passwords, filename listings, file content, as well as dates and times that files were accessed are recoverable from a personal computer which has interacted with the above cloud storage services [40-42]. An iPhone device was used to access the dataset through the 'on-device' web browser in order to interact with the specified iOS cloud storage application. A logical extraction using MicroSystemation's XRY was then performed. Their analysis of the logical extraction revealed that a number of artefacts can be recovered from various locations on the device which included the account services' username, as well as the filenames of viewed files [40-42]. However, it is worth noting that the content of the files stored in the cloud storage services were not recovered from the iPhone. Quick and Choo noted that future work should examine the physical acquisition of iPhone devices to determine if this method of acquisition can be used to recover the files stored in the cloud storage services [40-43].

Separately, Quick and Choo [45] investigated the modification of file content and metadata when potential evidence is downloaded and collected from a cloud storage account. Quick and Choo [45] reported that the cryptographic hashes calculated from file manipulations, like uploading, storing and downloading, using Dropbox, SkyDrive (now called OneDrive) and Google Drive reveal that no changes were made to the files' content. Quick and Choo [45] also state that after further analysing the downloaded files, notable changes were visible in the timestamp metadata. This was particularly evident in the 'last accessed' and 'file creation' time-stamps which indicated the last interaction with the cloud storage client software [45].

Martini and Choo [46] focused on client and server-side forensic investigations involving the 'ownCloud' service. Martini and Choo reported that forensic artefacts found on the client machine can link a user to a particular 'ownCloud' instance [46]. Furthermore, Martini and Choo recovered authentication and file metadata from the client, which were then used to decrypt files stored on the server [46]. Martini and Choo [47] have also examined XtreemFS, a distributed filesystem commonly used in cloud computing environments, and documented both client and server-side artefacts which may be relevant to forensic investigations.

As part of the "Cloud Computing and The Impact on Digital Forensic Investigations" (CLOIDIFIN) project, Biggs and Vidalis [48] reported that very few High Tech Crime Units in the United Kingdom were prepared to deal with crimes involving cloud computing. As a result, Biggs and Vidalis believe a 'cloud storm' will create difficulties and challenges for law enforcement investigators charged with investigating such environments [48]. Taylor, et al., [28] have extensively examined the legal issues surrounding cloud computing and comment that any evidence gathered from the cloud should be conducted within local laws and legislation. Phillips [49] discussed the issue of keeping a chain of custody for such an investigation and has argued that the cloud is a dynamic paradigm and physically isolating it to conduct an investigation could be a daunting task for the investigator.

When multiple devices are used to access data in the cloud, the issues with the dynamic paradigm are exacerbated. A prime example of this impediment is the increasing use of mobile devices such as smartphones to access data stored in a cloud [12, 38]. A smartphone device is distinguishable from a traditional mobile phone by its superior processing capabilities, a larger storage capacity, as well as its ability to run complex operating systems and applications [50]. From an evidentiary perspective, the smartphone potentially contains a considerable amount of forensic evidence. This potential is demonstrated in a study where researchers recovered more than 11,000 data artefact's from 49 predominately low-end devices [51]. As with a traditional mobile phone, the smartphone not only stores call logs, text messages and personal contacts, but it also has the ability to store web-browsing artefact's, email messages, GPS coordinates, as well as third-party application related data [52-54]. There are a number of tools that can be used to perform a data acquisition from a smartphone. Examples of these tools include Cellebrite's Universal Forensic Extraction Device (UFED) [55];

MicroSystemation's XRY tools [56]; The Mobile Internal Acquisition Tool [57]; Paraben's Device Seizure [58] and RAPI Tools [59]. These forensic toolkits make it possible to investigate mobile devices that have been used to access cloud storage providers and extract evidence without directly accessing the cloud storage provider's service. However, there is currently a lack of research as to the relationship between the residual data retained on multiple mobile devices subsequent to cloud interaction. There is also a lack of research into the impact that cloud storage applications have on mobile device residual data.

## 3. Experiment Design

The lack of research examining the gap between device residual data and existing cloud data prompted research into the following hypothesis:

**H$_1$:** Smartphone devices present a partial view of the data held in cloud storage services, which can be used as a proxy for evidence held on the cloud storage service itself.

**H$_2$:** The manipulation of different cloud storage applications influences the results of data collection from a smartphone device.

**H$_3$:** Different versions of cloud storage applications implemented on diverse operating systems retain varying amounts of residual data.

To address the hypotheses, the following questions were proposed:

1. To what extent can data stored in a cloud storage provider be recovered from a smartphone device that has accessed the service?
2. What features of the cloud application influence the ability to recover data stored in a cloud storage service from a smartphone device?
3. Do different versions of a cloud application used on the smartphone devices affect the ability to recover data stored in a cloud storage service from a smartphone device?
4. What metadata concerning the cloud storage service can be recovered from a smartphone device and what does the metadata data, recovered from a smartphone device, reveal about further files stored in the cloud service?
5. Does the amalgamation of files recovered from two or more versions of a specific cloud storage application provide a more complete dataset of files stored in the cloud service?

An experiment was devised to support the hypotheses and research questions proposed above. The experiment was broken into six stages. The six stages included: 1) preparing the smartphone device and installing the cloud application; 2) loading a dataset to a cloud storage provider; 3) connect to the data through the application on the smartphone; 4) performing various file manipulations to the dataset and smartphone device; 5) processing the device using the Universal Forensic Extraction Device (UFED); and 6) using a number of forensic tools to extract the files and artefacts from the resulting memory dumps.

The forensic tools used in this experiment were the Cellebrite Universal Forensic Extraction Device (UFED) version 1.8.5.0 and its associated application the 'Physical Analyzer' version 3.7.0.352; FTK Imager and FTK Toolkit version 4.0. The smartphone devices were processed with the UFED tools. The memory card used in the HTC Desire was processed using FTK Imager. The memory dumps were examined using a combination of Physical Analyzer and the FTK toolkit. Three smartphone devices were selected for use in this experiment: an Apple iPhone 3G and two HTC Desire devices. Table 1 – Smartphone Device Features highlights the notable features of these devices. These devices were selected for two reasons. First, they are compatible with the choice of forensic toolkit (UFED) used to perform a physical dump of the internal memory. Second, the operating systems used on these devices represent the two most popular smartphone operating systems in use [60]. Although more recent devices with newer versions of both the Android and iOS operating system exist, a lack of support from the forensic tools to perform a physical acquisition meant that these newer devices could not be included in the experiment. The decision to use these specific devices and tools was a pragmatic decision based on practicality and availability to the authors.

| Feature | iPhone 3G | HTC Desire | HTC Desire |
|---|---|---|---|
| **Operating system** | iOS v. 3 | Android v. 2.1 | Android v. 2.3 |
| **Internal memory** | 8 GB storage | 576 MB RAM | 576 MB RAM |
| **Memory card** | No | Yes (4 GB) | Yes (4 GB) |

**Table 1. Smartphone Device Features**

The selection criteria for the smartphone devices limited the number of cloud storage applications available to only the applications compatible with both operating systems. The scope of the experiment was limited in the following ways:

- This experiment was conducted in the United Kingdom, where Global System of Mobile (GSM) is the predominant mobile phone type, therefore non-GSM mobile devices were not considered;

- A number of smartphone devices which run either iOS or Android were not considered due to compatibility issues with the toolkit; and

- Various cloud storage applications were not considered because they do not support either or both of the chosen operating system platforms.

The original implementation of this experiment used an iPhone 3G running iOS version 3.0 and an HTC Desire running Android version 2.1. The cloud storage applications for iOS included: Dropbox v. 1.4.7, Box v. 2.7.1, SugarSync v. 3.0, and Syncplicity v. 1.6. The cloud storage applications for the Android device included: Dropbox v. 2.1.3, Box v. 1.6.7, SugarSync v. 3.6, and Syncplicity v. 1.7. The experiment was then extended and repeated using an HTC Desire running Android version 2.3. Newer versions of the Android cloud storage applications implemented in this portion of the experiment included: Dropbox v. 2.2.2, Box v. 2.0.2, SugarSync v. 3.6.2, and Syncplicity v. 2.1.1. Extending the experiment provides the opportunity to compare the results obtained between different versions of specific cloud storage applications. Updating the operating system and applications used on the iPhone device was considered. However, based on discontinued application support for iPhone 3G, a lack of support, at the time of the experiment, from the forensic tools for newer versions of the iPhone, as well as device availability, a pragmatic decision was made not to include an iOS device in the extended experiment.

A pre-defined dataset was created, which was comprised of 20 files, made up of image (JPEG), audio (MP3), video (MP4), and document (DOCX and PDF) file types. Table 2 – Experimental Dataset defines the files used in this dataset. The same dataset was used in both the original and extended experiments. The following steps were used in both the original and extended experiments. These steps were repeated every time the experiment was reset for a different cloud storage application.

1. The smartphone was 'hard reset', which involved restoring the default factory settings on the device. In the case of the HTC Desire, the SD memory card was forensically wiped using The Department of Defence Computer Forensics Lab tool – *dcfldd* [61]. These steps were done to remove any previous data stored on the devices and the memory card.

2. The device was then connected to a wireless network which was used to gain access to the Internet. The cloud storage application was downloaded and installed either via the Android or Apple 'app market', depending on the device used. The default installation and security parameters were used during the installation of the application.

3. The cloud storage application was executed, and a new user account was created using a predefined email address and a common password for that cloud storage application.

4. After the test account was created, the application was 'connected' to the cloud storage provider's services, which meant the device was now ready to receive the dataset.

5. A personal computer running Windows 7 was used to access the test account created in Step 4 and the dataset was then uploaded to the cloud storage provider using a web browser. The date and time the files were uploaded to the cloud storage provider was noted. The smartphone was synchronized with the cloud storage provider, to ensure the dataset was visible via the smartphone application.

6. When the entire dataset was visible on the smartphone, a number of manipulations were made to files in the dataset. Table 2 – Experimental Dataset summarizes these manipulations. These included:

    - a file being viewed or played;
    - a file viewed or played then saved for offline access;
    - a file viewed or played then deleted from the cloud storage provider; and
    - some files were neither opened/played nor deleted (no manipulation).

7. The smartphone and cloud storage application were also manipulated in one of the following ways:

    - *Active power state* - the smartphone was not powered down and the application's cache was not cleared;
    - *Cache cleared* - the applications cache was cleared;
    - *Powered off* - the smartphone was powered down; and
    - *Cache cleared and powered off* - the applications' cache was cleared and the smartphone was powered off.

These manipulations were done to mimic various scenarios a forensic investigator could encounter during an investigation. The smartphone was then removed from the wireless network to prevent any accidental modification to the dataset.

| Filename | Size (bytes) | Manipulation |
|---|---|---|
| **01.jpg** | 43183 | File viewed/played |
| **02.jpg** | 6265 | File viewed/played and saved for offline access |
| **03.jpg** | 102448 | No manipulation |
| **04.jpg** | 5548 | File viewed/played and then deleted |
| **05.mp3** | 3997696 | File viewed/played |
| **06.mp3** | 2703360 | File viewed/played and saved for offline access |
| **07.mp3** | 3512009 | No manipulation |
| **08.mp3** | 4266779 | File viewed/played and then deleted |
| **09.mp4** | 831687 | File viewed/played |
| **10.mp4** | 245779 | File viewed/played and saved for offline access |
| **11.mp4** | 11986533 | No manipulation |
| **12.mp4** | 21258947 | File viewed/played and then deleted |
| **13.pdf** | 1695706 | File viewed/played |
| **14.pdf** | 471999 | File viewed/played and saved for offline access |
| **15.pdf** | 2371383 | No manipulation |
| **16.pdf** | 1688736 | File viewed/played and then deleted |
| **17.docx** | 84272 | File viewed/played |
| **18.docx** | 85091 | File viewed/played and saved for offline access |
| **19.docx** | 14860 | No manipulation |
| **20.docx** | 20994 | File viewed/played and then deleted |

**Table 2. Experimental Dataset**

8. After the above manipulations, the smartphone device was processed to create a forensic dump of its internal memory. In the case of the HTC Desire, the Secure Digital (SD) memory card was processed separately from the smartphone. The HTC Desire was processed directly using the UFED, while a binary image of the SD card was created using FTK Imager. Before the HTC

Desire was processed, the USB debugging option was enabled on the smartphone. This is required by the UFED to create the binary images from the device. The default parameters for a Physical Extraction on the UFED were selected, and the make and model of the device were provided. In the case of the SD card, the default parameters were used to create a binary image of the storage card. The resulting binary images were saved to a forensically wiped 16 GB USB flash drive. The extraction process for the iPhone differed from that of the HTC Desire as the device was processed using the Physical Analyzer 'add-on', which is designed to extract binary images from the iPhone. A step-by-step wizard provided instructions on how to prepare the device for the extraction. From the selection menu, the User partition was selected for extraction from the device, and the resulting memory dump was saved to a 16 GB USB flash drive.

9. The images extracted from the smartphone device were then loaded into Physical Analyzer, where the iOS and Android file systems were reconstructed. FTK 4 was used as the primary tool for analysis. This involved extracting the partitions from the dumps in Physical Analyzer and then examining them using FTK. Analysis techniques used included: string searching for the password, filtering by file types and browsing the iOS and Android file systems.

## 4. Findings

A summary of files recovered from the devices is shown in the following tables: Table 3 – Dropbox Files Recovered, 4 – Box Files Recovered, 5 – SugarSync Files Recovered and 6 – Syncplicity Files Recovered. Several observations can be drawn from these results. Smartphone devices can be used to recover artefacts related to cloud storage services. These artefacts can include the files stored in the cloud storage service which have been accessed using the smartphone device and metadata associated to user and service activity. The exception to this was the recovery of a thumbnail of the JPEG image file not viewed on the device (03.jpg), which was recovered from ten of the twelve applications examined.

The chances of recovering a file increase if the file has been saved for offline viewing. Files which were marked for offline viewing were recovered from all the applications except from version 2.0.2 of the Box Android application. The results also indicate that different versions of the Android applications can result in different files being recovered from a smartphone device. This finding was particularly evident for the Box and Syncplicity applications. The two different versions of these applications resulted in different files from the dataset being recovered. The metadata recovered from the devices included SQLite databases, text-based transaction logs, JavaScript Object Notation (JSON) and XML files. These metadata artefacts contained information related to user activity; account-specific information such as email addresses and described which files are stored in the cloud storage service.

An analysis of the memory dumps revealed that forensic artefacts can be recovered from the smartphone devices and in the case of the Android devices, the SD memory card. The Android operating system allows files to be stored in either the device's internal storage memory or on an external memory card [52]. The iPhone does not have an external storage device and all artefacts recovered from the device were from the internal storage memory. The SD memory card used with the Android devices contained files which were either deleted by the user or deleted as a result of the cache being cleared. Clearing the application's cache has an adverse effect on the recovery of files. This is more evident on the iPhone, which does not contain an SD card. Powering down the smartphone devices did not have an effect on the recovery of data. As a result, the files recovered were identical to that of the active power state scenario.

Artefacts stored in the internal memory of the Android devices can be recovered from a sub-folder named after the application name. This sub-folder can be recovered from the path `/data/data` [52]. Unlike the internal storage device, applications can store data in any location on the SD memory card [62]. Therefore, the location of evidence on the SD card varies, depending on the application being investigated. The iOS file system creates a sub-folder for each installed application under the directory `/private/var/mobile/Applications` in the User partition [53]. The name of the application directory installed under this location is assigned a unique 32 character alphanumeric folder name [63]. The folder name is different for each application installed on the device. Artefacts related to the iOS applications were stored under this folder location.

## 4.1 Detailed Dropbox Findings

On the HTC Desire, the forensic toolkits recovered nine files from both Android versions of Dropbox. Depending on application and device manipulation, either five or seven files were recovered from the iOS version of Dropbox. The results of which files were recovered from the Dropbox application are shown in Table 3 – Dropbox Files Recovered.

| Filename | Android Application Version 2.1.3 | | | | Android Application Version 2.2.2 | | | | iOS Application Version 1.4.7 | | | |
|---|---|---|---|---|---|---|---|---|---|---|---|---|
| | APS | CC | PWD | CC & PWD | APS | CC | PWD | CC & PWD | APS | CC | PWD | CC & PWD |
| 01 | T | T | T | T | T | T | T | T | T | | T | |
| 02 | ✓ | ✓ | ✓ | ✓ | ✓ | ✓ | ✓ | ✓ | ✓ | ✓ | ✓ | ✓ |
| 03 | T | T | T | T | T | T | T | T | T | | T | |
| 04 | T | T | T | T | T | T | T | T | | | | |
| 05 | | | | | | | | | | | | |
| 06 | ✓ | ✓ | ✓ | ✓ | ✓ | ✓ | ✓ | ✓ | ✓ | ✓ | ✓ | ✓ |
| 07 | | | | | | | | | | | | |
| 08 | | | | | | | | | | | | |
| 09 | | | | | | | | | | | | |
| 10 | ✓ | ✓ | ✓ | ✓ | ✓ | ✓ | ✓ | ✓ | ✓ | ✓ | ✓ | ✓ |
| 11 | | | | | | | | | | | | |
| 12 | | | | | | | | | | | | |
| 13 | ✓ | D | ✓ | D | ✓ | D | ✓ | D | ✓ | | ✓ | |
| 14 | ✓ | ✓ | ✓ | ✓ | ✓ | ✓ | ✓ | ✓ | ✓ | ✓ | ✓ | ✓ |
| 15 | | | | | | | | | | | | |
| 16 | D | D | D | D | D | D | D | D | | | | |
| 17 | ✓ | D | ✓ | D | ✓ | D | ✓ | D | ✓ | | ✓ | |
| 18 | ✓ | ✓ | ✓ | ✓ | ✓ | ✓ | ✓ | ✓ | ✓ | ✓ | ✓ | ✓ |
| 19 | | | | | | | | | | | | |
| 20 | D | D | D | D | D | D | D | D | | | | |

**Table 3: Dropbox Files Recovered**

**APS = Active Power State; PWD = Powered Down; CC = Cache Cleared; CC & PWD = Cache Cleared and Powered Down; ✓ = File Recovered; D = Deleted File Recovered; T= Thumbnail Recovered**

### 4.1.1 Android Applications

Files stored in the Dropbox service can be recovered from two locations on the SD card. These locations and their contents are valid for both versions of the Android Dropbox application. First, thumbnails of the JPEG images were recovered from the path `/Android/data/com.dropbox.android/cache/thumbs/`. Second, files which were saved for offline viewing and the document files which were viewed and not deleted on the device were recovered from the path `/Android/data/com.dropbox.android/files/scratch`. Analyses of the 'unallocated space' for both Android applications revealed that the two document files which were deleted (16.pdf and 20.docx) were still physically stored on the SD card. These two document files were recovered by FTK.

Metadata related to both versions of the Dropbox application were recovered from the internal memory of the smartphone. The metadata recovered were valid for both versions of the Android application. This metadata consisted of two SQLite databases and a transaction log. The two databases were recovered from the path `/data/data/com.dropbox.android/databases/`. The first database, `db.db`, has a table called `dropbox`, which contains metadata related to the files currently stored in the service, i.e., files which have not been deleted from the Dropbox service. Fields identified from this table are shown in Appendix A.

The second database, `prefs.db`, has a table called `DropboxAccountPrefs` which contains metadata related to the end-user. Information hich can be recovered includes the user's name and email address used to register for the Dropbox service. A transactional log called `log.txt` is created by the Dropbox application to record service and user-related events including: the creation of a new user account; successful and unsuccessful login attempts; as well as which files are synchronized to the Dropbox service. A UNIX epoch timestamp accompanies the documented event. This log can be recovered from the path `data/data/com.dropbox.android/files`.

Clearing the cache of both versions of the Dropbox Android application, removes the documents viewed and not deleted using the smartphone device (13.pdf and 17.docx), which are stored in the `com.dropbox.android/files/scratch` directory. These files were still physically stored on the SD card and are recovered by FTK. The files saved for offline access, JPEG thumbnails and metadata remain unaltered.

### 4.1.2 iOS Application

On the iOS device, a number of files stored in the Dropbox service were recovered from a sub-folder called `Dropbox`, which can be found in the path `/Library/Caches`. The following files were recovered from this location: thumbnails of three JPEG images (01.jpg, 02.jpg and 03.jpg); five files saved for offline access; and PDF and DOCX files, viewed but not deleted from the device (13.pdf and 17.docx). No other files stored in the Dropbox service were recovered from the iOS device.

The metadata artefacts recovered include an SQLite database, property list (plist) files and text-based logs. These metadata artefacts described user activity and the files stored in the service. The main metadata repository is an SQLite database called `Dropbox.sqlite` which contains metadata about the files stored in the Dropbox service. This database can be recovered from a sub-folder called `/Documents` from the application's root directory. The `ZCACHEDFILE` table within this database contains metadata related to the files recovered from the directory located at `/Library/Caches/Dropbox`. Fields identified from the `ZCACHEDFILE` are shown in Appendix A.

Additional metadata related to the files which were saved as 'favorite' and user-specific information can be recovered from two property list (plist) files. The first plist file located from the path `/Library/Preferences/com.getdropbox.Dropbox.plist` contains the email address used for the Dropbox account and information related to files which were saved as 'favorite'. The second plist file called `FavoriteFiles.plist` located from the path `/Library/Caches/` contains further information about files which were downloaded and saved as 'favorite'. Metadata which can be recovered from the `FavoriteFiles.plist` file includes the size of the file in bytes, the last modified time, the file name and if the file has been deleted.

Two transaction logs were also recovered from the iOS device. The first log called `Analytics.log` records user-related activity and can be recovered from the path `/Library/Caches`. Each entry in the log is accompanied by a UNIX epoch timestamp. Figure 1 shows an example record from the `Analytics.log` file, which describes a PDF file which was viewed and then saved for offline access. The second log, `run.log`, which can be recovered from the path `/tmp/` contains additional information about service-related transactions performed by Dropbox. When the Dropbox iOS application cache is cleared on the device, the only files which remain are those five files saved for offline access. This action also affects the `Dropbox.sqlite` database. When the application's cache is cleared the database only contains metadata for the five files which remain on the device. All other metadata artefacts remain unchanged.

```
{"retry":0,"favorite":false,"extension":"pdf","id":23,"cached":false,"ts":"1335445641.29","event":"file.view.start","size":1695706}
{"id":23,"ts":"1335445641.31","size":1695706,"event":"download.start","extension":"pdf","connection":"wifi"}
{"ts":"1335445641.84","screen":"DocumentViewController","event":"screen.view"}
{"id":23,"ts":"1335445657.75","size":1695706,"event":"download.success","extension":"pdf"}
{"id":23,"event":"file.view.success","ts":"1335445659.92"}
{"ts":"1335445669.71","screen":"SearchableFolderListController","event":"screen.view"}
{"ts":"1335445670.04","cached":true,"path_hash":912,"event":"metadata.load.start"}
{"path_hash":912,"event":"metadata.load.unchanged","ts":"1335445673.07"}
```

**Figure 1: Analytics.log file describing a PDF file which was viewed and then saved for offline access using the Dropbox iOS application**

## 4.2 Detailed Box Findings

From the Android applications, the forensic toolkits recovered fifteen files from version 1.6.7 and between four and six from version 2.0.2, depending on application and device manipulation. Five files were recovered from the iOS version of the Box application. The files which were recovered from the Box application are summarized in Table 4 – Box Files Recovered.

### 4.2.1 Android Applications

On the Android devices, Box-related artefacts varied between the two versions of the application. Artefacts related to version 1.6.7 of the Box application were recovered from three locations on the SD card. The files saved for offline access (02.jpg, 06.mp3, 10.mp4, 14.pdf and18.docx) were recovered from the path `/Box/email_address/`, where `email_address` is the email address used to register for the service. This version of the application caches any files which have been viewed on the device. These can be recovered from the directory `/Android/data/com.box.android/cache/filecache`. Fifteen files from the dataset were found in this directory. The files missing are those which are marked as 'no manipulation' in Table 2 – Dataset. Thumbnails of all four JPEG images (01-04.jpg) can be recovered from a sub-folder of the above location called `/tempfiles/box_tmp_images`.

Artefacts related to version 2.0.2 of the Box application were recovered from four locations on the SD card. This version of the Box application encrypts the cache folders used by the service. Three encrypted folders called `dl_cache`, `dl_offline` and `previews` were recovered from the path `/Android/data/com.box.android/cache`. No files from the dataset were recovered from these three folders. Thumbnails of all four JPEG images can be recovered from the path `/data/data/com.box.android/cache/tempfiles/box_tmp_images`. The six MP3 (05.mp3, 06.mp3 and 08.mp3) and MP4 (09.mp4, 10.mp4 and 12.mp4) files viewed on the device can be recovered from a sub-folder called `working` located under the path `/data/data/com.box.android/cache`. This version of the Box application creates an additional folder of interest called `previews` which can be recovered from the path `/data/data/com.box.android/files/`. The `previews` folder contains PNG image files of

'snapshots' of the text-based documents (DOCX and PDF) and JPEG images from the dataset which have been viewed using the device.

The metadata artefacts for both versions of the Box application can be recovered from the smartphone, which unless stated were the same for both versions of the application. The Box application creates a JavaScript Object Notation (JSON) file called `json_static_model_emailaddress_0`, where `emailaddress` is the email address used to sign-up to the Box service. This file can be recovered from the path `/data/data/com.box.android/files/`. This JSON file contains property metadata about the files stored in this Box service. Fields identified from the JSON file are described in Appendix B.

| Filename | Android Application Version 1.6.7 | | | | Android Application Version 2.0.2 | | | | iOS Application Version 2.7.1 | | | |
| --- | --- | --- | --- | --- | --- | --- | --- | --- | --- | --- | --- | --- |
| | APS | CC | PWD | CC & PWD | APS | CC | PWD | CC & PWD | APS | CC | PWD | CC & PWD |
| 01 | ✓ | D | ✓ | D | T | T | T | T | T | T | T | T |
| 02 | ✓ | D | ✓ | D | T | T | T | T | ✓ | ✓ | ✓ | ✓ |
| 03 | T | T | T | T | T | T | T | T | T | T | T | T |
| 04 | ✓ | D | ✓ | D | T | T | T | T | T | T | T | T |
| 05 | ✓ | D | ✓ | D | ✓ | | ✓ | | | | | |
| 06 | ✓ | D | ✓ | D | ✓ | | ✓ | | ✓ | ✓ | ✓ | ✓ |
| 07 | | | | | | | | | | | | |
| 08 | ✓ | D | ✓ | D | ✓ | D | ✓ | D | | | | |
| 09 | ✓ | D | ✓ | D | ✓ | D | ✓ | D | | | | |
| 10 | ✓ | D | ✓ | D | ✓ | D | ✓ | D | ✓ | ✓ | ✓ | ✓ |
| 11 | | | | | | | | | | | | |
| 12 | ✓ | D | ✓ | D | ✓ | D | ✓ | D | | | | |
| 13 | ✓ | D | ✓ | D | | | | | | | | |
| 14 | ✓ | D | ✓ | D | | | | | ✓ | ✓ | ✓ | ✓ |
| 15 | | | | | | | | | | | | |
| 16 | ✓ | D | ✓ | D | | | | | | | | |
| 17 | ✓ | D | ✓ | D | | | | | | | | |
| 18 | ✓ | D | ✓ | D | | | | | ✓ | ✓ | ✓ | ✓ |
| 19 | | | | | | | | | | | | |
| 20 | ✓ | D | ✓ | D | | | | | | | | |

**Table 4: Box Files Recovered**

**APS = Active Power State; PWD = Powered Down; CC = Cache Cleared; CC & PWD = Cache Cleared and Powered Down; ✓ = File Recovered; D = Deleted File Recovered; T= Thumbnail Recovered**

A second location containing metadata related to the Box application can be found under the path `/data/data/com.box.android/shared_prefs`. This location contains a number of XML files which contain user and service-specific information. The files created in this location vary between the two versions of the application. The files recovered from this path include:

- *myPreference.xml* - which contains the authentication token associated with this particular account and the email address used to register for the Box service. This can be recovered from both versions of the application.
- *Preview_Num_Pages.xml* – this file is related to the folder recovered from the path `/data/data/com.box.android/files/previews` and contains metadata such as the `mId` of the file whose 'preview' is stored in the folder as well as the number of 'preview' pages. This file can only be recovered from version 2.0.2 of the application.
- *Downloaded_Files.xml* - contains metadata about the files downloaded to the SD card from the Box service. `Long name` is the ID number assigned to that particular file and `value` is the date and time the file was deleted. The data and time is stored as a UNIX epoch timestamp. This file name is only valid for version 1.6.7 of the Box application. The file is renamed to `offlineFileSharedPreferences.xml` for version 2.0.2 but contains the same metadata related to files saved for offline viewing.

When the cache is cleared on the version 1.6.7 of the Box application, the contents of the `Android/data/com.box.android/cache/filecache` and the `/Box/email_address/` directories are deleted and recovered by FTK. All other files and metadata related to the Box service are not affected. When the cache is cleared on version 2.0.2 of the application, the three encrypted folders, the files stored in the `/data/data/com.box.android/cache` and `working` folders as well as the PNG 'snapshot' files stored on the device are deleted but can be recovered using FTK. All other files and metadata related to the Box service are not affected.

### 4.2.2 iOS Application

On the iOS device, the files and metadata related to the Box service can be recovered from three main locations under the application's root directory. The files saved for offline viewing (02.jpg, 06.mp3, 10.mp4, 14.pdf and 18.docx) can be recovered from a sub-folder located under the path `/Documents/SavedFiles`. The thumbnails of the four JPEG images stored in the Box service can be found in the sub-folder `/Library/Caches/Thumbnails`. No other files from the dataset were recovered from the Box service. Metadata related to files stored in the service can be recovered from a SQLite database called `BoxCoreDataStore.sqlite` found under the sub-folder `/Documents/`. This database contains a table called `ZBOXBASECOREDATA`, which includes property metadata for all twenty files in the dataset. The metadata which can be recovered from this database is described in more detail in Appendix B. Additional information which can be recovered from this database includes the username and email address used to create the Box account and a unique authentication token assigned to the user account. Clearing the cache of the iOS Box application has no effect on the data or metadata stored on the device.

## 4.3 Detailed SugarSync Findings

On the HTC Desire, the forensic toolkits recovered eleven files from both Android versions of SugarSync and, depending on application and device manipulation, either seven or fifteen were recovered from the iOS version of the application. The results of which files were recovered from the SugarSync application are shown in Table 5 – SugarSync Files Recovered.

### 4.3.1 Android Applications

Files stored in the SugarSync service can be recovered from three locations on the SD card. These locations and their contents are valid for both versions of the Android SugarSync application. First, the three PDF files viewed on the smartphone (13.pdf, 14.pdf and 16.pdf) can be recovered from a folder called `/.sugarsync` located in the root directory of the application. Second, the thumbnails of all four JPEG images, the three JPEG images viewed on the device (01.jpg, 02.jpg and 04.jpg) and four document files viewed on the device (13.pdf, 16.pdf, 17.docx and 20.docx) can be recovered from a sub-folder called `/.httpfilecache` found in the above location. The third location is a folder called `/MySugarSyncFolders` located in the root directory of the application where the five files saved for offline viewing (02.jpg, 06.mp3, 10.mp4, 14.pdf and18.docx) can be found.

Metadata related to the SugarSync service can be recovered from two text-based transaction logs and an SQLite database. The metadata artefacts recovered from the device are valid for both versions of the Android application. The first transaction log is called `sc_appdata` and is recovered from the path `/data/data/com.sharpcast.sugarsync/app_SugarSync/SugarSync/`. This log contains the user's email address used to register for the service, the unique ID assigned to the user and a hash of the user's password.

| Filename | Android Application Version 3.6 | | | | Android Application Version 3.6.2 | | | | iOS Application Version 3.0 | | | |
|---|---|---|---|---|---|---|---|---|---|---|---|---|
| | APS | CC | PWD | CC & PWD | APS | CC | PWD | CC & PWD | APS | CC | PWD | CC & PWD |
| 01 | ✓ | D | ✓ | D | ✓ | D | ✓ | D | ✓ | | ✓ | |
| 02 | ✓ | ✓ | ✓ | ✓ | ✓ | ✓ | ✓ | ✓ | ✓ | ✓ | ✓ | ✓ |
| 03 | T | T | T | T | T | T | T | T | | | | |
| 04 | ✓ | D | ✓ | D | ✓ | D | ✓ | D | ✓ | | ✓ | |
| 05 | | | | | | | | | ✓ | ✓ | ✓ | ✓ |
| 06 | ✓ | ✓ | ✓ | ✓ | ✓ | ✓ | ✓ | ✓ | ✓ | ✓ | ✓ | ✓ |
| 07 | | | | | | | | | | | | |
| 08 | | | | | | | | | ✓ | ✓ | ✓ | ✓ |
| 09 | | | | | | | | | ✓ | | ✓ | |
| 10 | ✓ | ✓ | ✓ | ✓ | ✓ | ✓ | ✓ | ✓ | ✓ | ✓ | ✓ | ✓ |
| 11 | | | | | | | | | | | | |
| 12 | | | | | | | | | ✓ | | ✓ | |
| 13 | ✓ | ✓ | ✓ | ✓ | ✓ | ✓ | ✓ | ✓ | ✓ | | ✓ | |
| 14 | ✓ | ✓ | ✓ | ✓ | ✓ | ✓ | ✓ | ✓ | ✓ | ✓ | ✓ | ✓ |
| 15 | | | | | | | | | | | | |
| 16 | ✓ | ✓ | ✓ | ✓ | ✓ | ✓ | ✓ | ✓ | ✓ | | ✓ | |
| 17 | ✓ | D | ✓ | D | ✓ | D | ✓ | D | ✓ | | ✓ | |
| 18 | ✓ | ✓ | ✓ | ✓ | ✓ | ✓ | ✓ | ✓ | ✓ | ✓ | ✓ | ✓ |
| 19 | | | | | | | | | | | | |
| 20 | ✓ | D | ✓ | D | ✓ | D | ✓ | D | ✓ | | ✓ | |

**Table 5: SugarSync Files Recovered**

**APS = Active Power State; PWD = Powered Down; CC = Cache Cleared; CC & PWD = Cache Cleared and Powered Down; ✓ = File Recovered; D = Deleted File Recovered; T= Thumbnail Recovered**

The second transaction log is called `sugarsync.log` and is recovered from the path `/data/data/com.sharpcast.sugarsync/app_SugarSync/SugarSync/log`. This log contains events related to the SugarSync service. For example, entries in this log include: the user authenticating with the service, the user downloading files on the device and an MP4 file being 'synced' from the service and then played on the device.

The SQLite database relevant to this application is called `SugarSyncDB` and can be recovered from the path `/com.sharpcast.sugarsync/databases`. This database has a table called `rec_to_offline_file_X`, where *X* is the unique ID number assigned to the user. This table contains metadata related to files saved for offline viewing and a UNIX epoch timestamp of when the file was saved for offline viewing.

When the cache is cleared on both versions of the Android application, the files affected are those stored under the location `/.sugarsync/.httpfilecache`, which are deleted from the SD card and recovered by FTK. All other files and metadata artefacts are not affected.

### 4.3.2 iOS Application

Files and metadata related to the iOS version of the SugarSync application can be recovered from four locations on the device under the application's root directory. The SugarSync service caches the files viewed on the device in a folder called `/tmp`. Files from the dataset can be recovered in two sub-folders within this location. The JPEG, MP4, DOCX and PDF files viewed on the device can be recovered from a sub-folder from the path `/tmp/http_cache`. The three MP3 files (05.mp3, 06.mp3 and 08.mp3) viewed on the device were recovered from the path `/tmp/cache`. The files which were saved for offline viewing (02.jpg, 06.mp3, 10.mp4, 14.pdf and18.docx) can be recovered from a sub-folder called `/MyiPhone` located under the `/Documents` directory.

The SugarSync service creates two main artefacts containing metadata related to the user and files stored in the service. These two artefacts can be recovered from the `/Documents` sub-folder. Account-specific information such as the email address used to register for the service can be recovered from a file called `ringo.appdata`. An SQLite database called `Ringo.sqlite`, contains a table called `ZSYNCOBJECT`. This table can be used to recover metadata related to the files saved for offline access. When the SugarSync application cache is cleared, the contents of the `/http_cache` folder are deleted. No other files and artefacts are affected when the cache is cleared.

## 4.4 Detailed Syncplicity Findings

From the Android Syncplicity applications, the forensic toolkits recovered nine files from version 1.7 and fifteen from version 2.1.1. Depending on application and device manipulation, either zero or fourteen were recovered from the iOS version of the application. The results of which files were recovered from the Syncplicity application are shown in Table 6 – Syncplicity Files Recovered.

### 4.4.1 Android Applications

On the Android devices, Syncplicity-related artefacts varied between the two versions of the application. For version 1.7 of the application, files were recovered from three locations on the SD card. Thumbnails of all four JPEG images can be recovered in the path `Android/data/com.syncplicity.android/cache/cachefu/image_cache`. The files which were saved for offline viewing (02.jpg, 06.mp3, 10.mp4, 14.pdf and 18.docx) can be recovered from a folder called `/Syncplicity`, which is stored in the root directory of the application. Version 1.7 of the application encrypts the cache folder used by the application. This folder can be recovered from the path `/Android/data/com.syncplicity.android/cache/private_syncp_file_cache_v3/encrypted/X`, where *X* is the unique ID assigned to the user. No files from the dataset were recovered from this location.

Further, files and metadata related to version 1.7 were recovered from four locations on the smartphone device. Files from the dataset can be recovered from the device in a directory located at the path `/data/data/com.syncplicity.android/files`. The files recovered from this location have been deleted, however, FTK was used to recover specific files from the dataset. The files recovered were the three JPEG files (01.jpg, 02.jpg and 04.jpg) and the three DOCX files

(17.docx, 18.docx and 20.docx) viewed and not deleted on the smartphone device. No other files were recovered from this location.

| Filename | Android Application Version 1.7 | | | | Android Application Version 2.1.1 | | | | iOS Application Version 1.6 | | | |
|---|---|---|---|---|---|---|---|---|---|---|---|---|
| | APS | CC | PWD | CC & PWD | APS | CC | PWD | CC & PWD | APS | CC | PWD | CC & PWD |
| 01 | D | D | D | D | ✓ | D | ✓ | D | ✓ | | ✓ | |
| 02 | ✓ | ✓ | ✓ | ✓ | ✓ | ✓ | ✓ | ✓ | ✓ | | ✓ | |
| 03 | T | T | T | T | T | T | T | T | | | | |
| 04 | D | D | D | D | ✓ | D | ✓ | D | ✓ | | ✓ | |
| 05 | | | | | ✓ | D | ✓ | D | ✓ | | ✓ | |
| 06 | ✓ | ✓ | ✓ | ✓ | ✓ | ✓ | ✓ | ✓ | ✓ | | ✓ | |
| 07 | | | | | | | | | | | | |
| 08 | | | | | ✓ | D | ✓ | D | ✓ | | ✓ | |
| 09 | | | | | ✓ | D | ✓ | D | ✓ | | ✓ | |
| 10 | ✓ | ✓ | ✓ | ✓ | ✓ | ✓ | ✓ | ✓ | ✓ | | ✓ | |
| 11 | | | | | | | | | | | | |
| 12 | | | | | ✓ | D | ✓ | D | | | | |
| 13 | | | | | ✓ | D | ✓ | D | ✓ | | ✓ | |
| 14 | ✓ | ✓ | ✓ | ✓ | ✓ | ✓ | ✓ | ✓ | ✓ | | ✓ | |
| 15 | | | | | | | | | | | | |
| 16 | | | | | ✓ | D | ✓ | D | ✓ | | ✓ | |
| 17 | D | D | D | D | ✓ | D | ✓ | D | ✓ | | ✓ | |
| 18 | ✓ | ✓ | ✓ | ✓ | ✓ | ✓ | ✓ | ✓ | ✓ | | ✓ | |
| 19 | | | | | | | | | | | | |
| 20 | D | D | D | D | ✓ | D | ✓ | D | ✓ | | ✓ | |

**Table 6: Syncplicity Files Recovered**

**APS = Active Power State; PWD = Powered Down; CC = Cache Cleared; CC & PWD = Cache Cleared and Powered Down; ✓ = File Recovered; D = Deleted File Recovered; T= Thumbnail Recovered**

Metadata artefacts related to version 1.7 which were recovered from the device included a text-based log, XML files and an SQLite database. These artefacts contained metadata related to both user activity and the files stored in the service. A text-based transaction log called `0000000000000000000.log.gz.tmp` contains metadata about the application and its interaction with the cloud service. This log can be recovered from the path `/data/data/com.syncplicity.android/app_log_syncplicity`. An SQLite database called `CacheDatabase`, can be recovered from the path `/data/data/com.syncplicity.android/databases`. This database contains a table called `Files`, which can be used to recover property metadata for all twenty files stored in the Syncplicity service. The information which can be recovered from the `Files` table can be seen in Appendix C.

A final source of metadata related to version 1.7 can be recovered from the path `/data/data/com.syncplicity/shared_prefs`. This folder contains the following XML files created by the application:

- *auth_prefs.xml* – this file contains the email address used to register for the Syncplicity service;

- *file_cache_preferences(X).deleted.xml* – X is a number and twenty-five different files can be found with such a naming convention. The format of the file is shown in Figure 2 below. These XML files can be used as a mapping for the encrypted directory found on the SD memory card.

```xml
<?xml version="1.0" encoding="utf-8" standalone="yes" ?>
- <map>
    <long name="FILE_CACHE_PREFERENCES_LAST_DECRYPTED_FILE_VERSION_ID" value="145789448" />
    <string name="FILE_CACHE_PREFERENCES_LAST_DECRYPTED_FILE_NAME">016.pdf</string>
  </map>
```

**Figure 2: Example of file_cache_preferences(X).deleted.xml file mapping retrieved from the Syncplicity Android application**

When the cache is cleared on version 1.7 of the application, the contents of the `cache/cachefu/image_cache` and encrypted folders are both deleted and recovered by FTK. No other files or artefacts are affected by the cache being cleared.

For version 2.1.1 of the application, files from the dataset were again recovered from both the SD card. Thumbnails of all four JPEG images can be recovered from the directory `/Android/data/com.syncplicity.android/cache/cachefu/image_cache`. As with the previous version, version 2.1.1 of the application also encrypts the cache folder used by the Syncplicity service. This folder can be recovered from the path `/Android/data/com.syncplicity.android/encrypted_ storage`. No files from the dataset were recovered from this location. This version of the Syncplicity application also contains a 'decrypted' cache folder where fifteen files from the dataset which were viewed on the device can be found. This folder can be recovered from the path `/Android/data/com.syncplicity.android/temporary_decrypted_storage`.

The metadata artefacts recovered from version 2.1.1 of the application included an SQLite database, a text-based transaction log and XML files. These artefacts were recovered from the internal memory of the device. A text transaction log called `00000000000000000.log.gz.tmp` contains metadata about the application and its interaction with the cloud service. This log can be recovered from the path `/data/data/com.syncplicity.android/app_log_syncplicity`. An SQLite database called `VIRTUAL_FILE_SYSTEM.db` was recovered from the path `/data/data/com.syncplicity.android/databases`. This database contains two tables of interest. The first table is called `Files` and contains metadata about all twenty files stored in the Syncplicity service. Appendix C shows the metadata which can be extracted from this table. The second table of interest is called `Files_and_Folders_to_Synchronize` which contains the names of the files saved for offline viewing. The XML files recovered from version 2.1.1 of the application can be found in the same location as the files in version 1.7: `/data/data/com.syncplicity/shared_prefs`. The contents of the XML files recovered from this location are the same for those recovered from version 1.7.

When the cache is cleared on version 2.1.1 of the application, the contents of the `/cache/cachefu/image_cache` and encrypted folders are both deleted and recovered by FTK. The files recovered from the `temporary_decrypted_storage` folder are also affected when the cache is cleared. This folder now contains only the files which were saved for offline access. All other files have been deleted and recovered by FTK. The `VIRTUAL_FILE_SYSTEM` database is also affected when the cache is cleared. No other files and artefacts are affected by the cache being cleared.

#### 4.4.2 iOS Application

Files and metadata related to the iOS version of the Syncplicity application can be recovered from four locations on the device under the application's root directory. The only location where files from the dataset can be recovered from the iOS device is a cache folder created by the application under the path `/Documents`. Fourteen out of the fifteen files viewed on the device can be recovered from this folder. The MP4 file viewed and then deleted (12.mp4) was the only file viewed on the device and not recovered from this location.

Metadata related to the iOS application consists of an SQLite database, a plist file and a text-based log. The SQLite database is called `syncplicity.sqlite` and can be recovered from the path `/Documents/`. This database contains a table called `ZFILES` which contains metadata about eighteen files from the dataset; the entries which are missing from this table are related to files 04.jpg and 08.mp3. The property metadata, which can be recovered from this table, is shown in Appendix C. Metadata related to the user account can be recovered from a plist file called `syncplicity.plist`, which can be found in the path `/library/preferences/com.syncplicity.ios`. This plist file can be used to recover information such as the type of account used in the service (free or paid) along with the first and last name of the user who registered for the account. The final location of metadata related to the iOS Syncplicity application is a transaction log called `syncplicity_0.log`, which can be found in the location `/library/caches`. This log contains user and service related transactions including files that were downloaded to the device and authentication token synchronization between the device and the Syncplicity service.

When the Syncplicity iOS application cache is cleared, the contents of the `/Documents` folder are deleted and no files are recovered from this location. No other files and artefacts are affected when the cache is cleared.

## 5. Discussion

The results described in the previous section can be used to provide answers to the research questions proposed in Section One. Forensic toolkits, including the Cellebrite UFED, can be used to recover data from a smartphone device that has accessed a cloud storage service. The proposed use of forensic and analysis toolkits currently available to the forensic community provides a practical solution for investigating cloud computing environments. The lack of forensic tools is commonly cited as a mainstream challenge for investigating cloud environments [2, 28, 34]. The results from this research suggest that end-devices, such as a smartphone, may contain evidence in relation to cloud storage services which may be important in an investigation, and that this resource should be considered and examined. Furthermore, the tools and methods used in the experiment to recover data from the smartphone device are widely used and accepted by the forensic community. It must be acknowledged that the files recovered from the smartphone devices present a 'snapshot in time' of the dataset stored in the cloud storage service. A file which is recovered from a smartphone does not mean that the file still exists in the cloud storage service, but provides an indication that at a point in time this file was stored in the service.

The results indicate that it is possible to recover files, providing a snap-shot in time, that indicates the existence of potential data that is stored in cloud services like Dropbox, Box, SugarSync and Syncplicity. On the HTC Desire, both deleted and available files were recovered. No deleted files were recovered from the iPhone. Certain file types were recovered more than other types. For example, the results show that JPEG thumbnail images were produced on all the devices running the Android applications. Thumbnail images were also recovered from the Dropbox and Box applications on the iOS device. In general, very few MP3 and MP4 files were recovered from all three devices. It is also interesting to note that more deleted files were recovered from the Box and Syncplicity applications than Dropbox or SugarSync applications on the HTC Desire.

The recovery of files from a smartphone device is affected by the user's manipulation of the device and the cloud storage application. The Box iOS application and version 1.7 of the Android Syncplicity application were the only applications where there was no difference in the number of files recovered from the 'active power state' and the 'cache cleared state'. The results also show that when the cache was cleared in all other instances, fewer files were recovered than from the 'active power state'. In the case of the iOS Syncplicity application, no files were recovered when the application's cache was cleared. User actions on specific files have shown to influence the recovery of these files. For example if a file has been viewed using the smartphone there is the opportunity for it to be recovered using forensic toolkits. This is provided that the user has not deleted the file, or cleared the application's cache. Files saved for offline access by the user can be recovered from the Android and iOS applications. There were two exceptions to recovering these files. The first is when the cache was cleared for the Syncplicity iOS application, none of the files saved for offline viewing were recovered. The second is when the cache was cleared for version 2.0.2 of the Android Box application, none of the files for offline viewing were recovered from any of the states. Deleted files were recovered from the Android devices. The recovery of these files is dependent upon them not being overwritten by new data on the SD memory card. No deleted files were recovered from the iOS device.

It is interesting to note that there are discrepancies in the number of files recovered depending on the version of the cloud storage service implemented on different mobile platforms and operating systems. While the versions of Dropbox produced the same number of files, there were vast differences between specific versions of Box, SugarSync and Syncplicity. A summary of the total number of active power state files recovered from the Android and iOS devices by cloud application is shown in Figure 3. This table presents the files recovered from the active power state excluding thumbnails and deleted files.

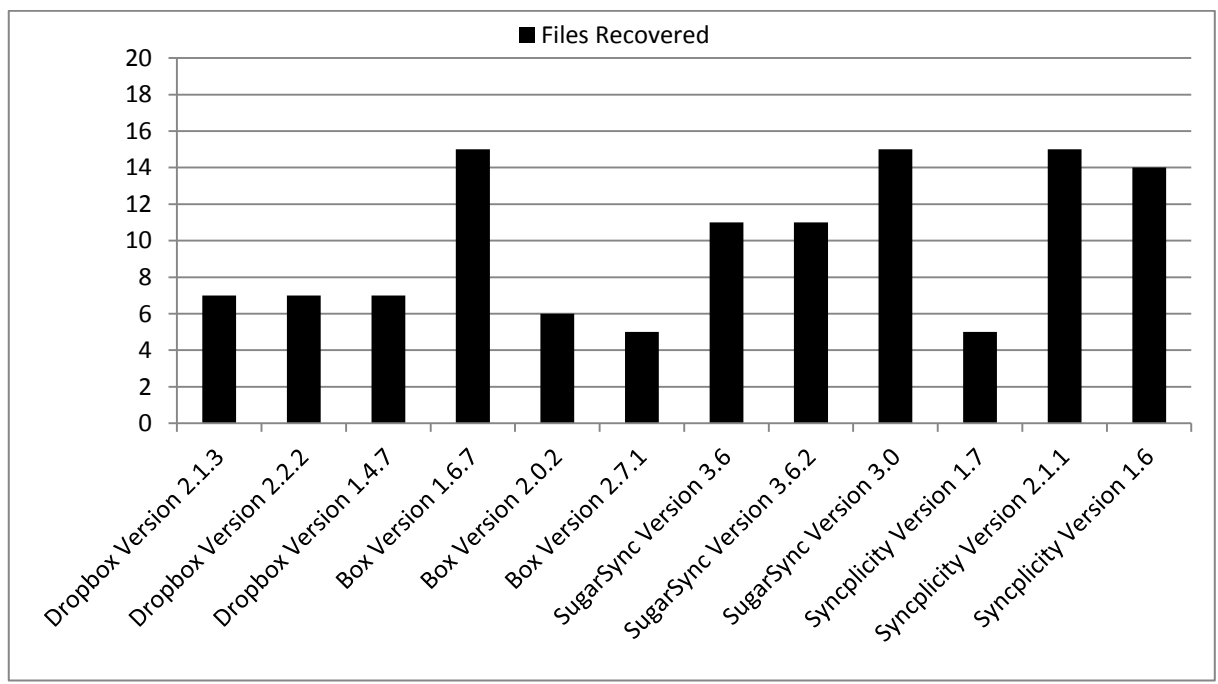

**Figure 3 : Total number of active power state recovered files**

Metadata was recovered from all the applications on all three devices. The metadata recovered included text-based transaction logs containing user and service activity, SQLite databases and JSON files containing property metadata data related to the files in the service, as well as XML files containing user-specific metadata such as login credentials. The metadata recovered from the devices can also present the investigator with a greater representation of the dataset stored in the cloud. For example, depending on the operating system platform, device and application manipulation, between four and fifteen files were recovered from the iOS and Android Box applications. However, the metadata artefacts recovered from these applications run on the Android and iOS devices, revealed information about files stored in the Box service which were not recovered from the device. The JSON files and SQLite databases recovered from the internal memory of these devices contained records for all twenty files stored in the Box service. The information which can be recovered includes the file names and unique identification number assigned to each file as well as user-specific identification numbers and email addresses used to register for the storage services. This metadata could help an investigator justify requesting a court order or warrant for a cloud storage provider to recover further files from the account being investigated [38].

Furthermore, using metadata artefacts recovered from the Box application, it is possible to download further files from the Box service. This can include files which were not recovered from the smartphone device itself. This information can be recovered from all three versions of the Box application which were included in this experiment. This is possible by constructing a direct link to the file stored in the Box service using the Box API [64]. This direct link requires three pieces of information from the smartphone device, for example from an Android device:

1. The authentication token, which can be recovered from the `myPreference.xml` file found in the path `/data/data/com.box.android/shared_prefs`. For example, in Figure 4, the authentication token is shown as: `:<string name="authToken">u5es7xli4xejrh89kr6xu14tks6grjn3</string>`;

2. The unique file ID number called `mId`, which is the ID number assigned to each file stored in the service. This information can be recovered from the `json_static_model_emailaddress_0` file stored in the directory `/data/data/com.box.android/files/`. The investigator requires the ID number for each file they wish to download from the Box service (Figure 5); and

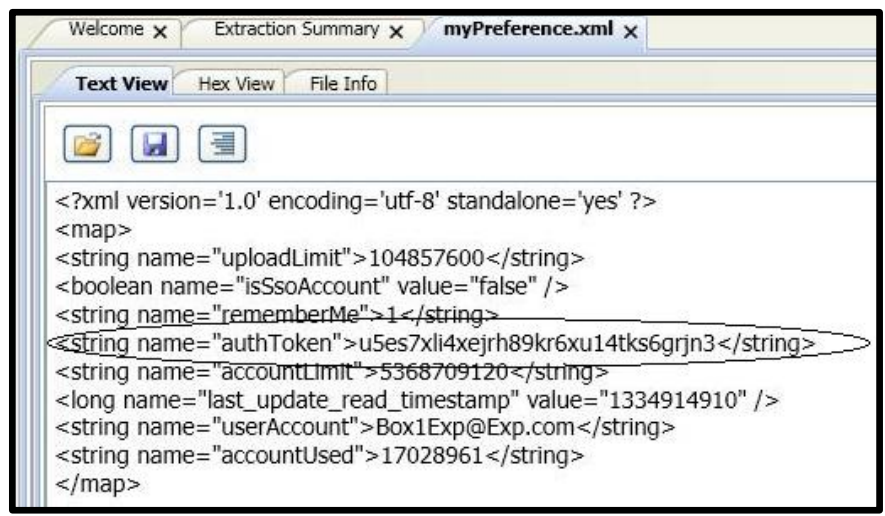

**Figure 4 : Metadata artefact containing the authentication token from Box Service**

3. A URL from the Box API [64]: `https://www.box.net/api/1.0/download/auth_token/file_id`, where auth_token is the authentication token for the account and file_id is the `mId` number of the file to be downloaded.

| mFileName | String | 003.jpg |
| mSmallThumbnail | String | https://mobile-api.box.com/api/data/bf1008/public/120120222/sma20/6e3b910225f37785bde9dad0f00f796e.gif |
| mLargeThumbnail | String | https://mobile-api.box.com/api/data/bf1008/public/120120222/larg20/6e3b910225f37785bde9dad0f00f796e.jpg |
| mLargerThumbnail | String | https://mobile-api.box.com/api/data/bf1008/public/120120222/larg20/6e3b910225f37785bde9dad0f00f796e.jpg |
| mPermissions | String | gdcenopstuvh |
| mPreviewThumbnail | String | https://mobile-api.box.com/api/data/bf1008/public/120120222/pre20/6e3b910225f37785bde9dad0f00f796e.jpg |
| mSha1 | String | 50f27a33962f80783eef3194c0275af77a8d4acf |
| mUpdated | Integer | 1334914771 |
| mId | Integer | 2072716499 |
| mSize | Integer | 102448 |
| mFolderId | Integer | 0 |
| mCreated | Integer | 1334914769 |
| mCommentCount | Integer | 0 |

**Figure 5 : mID value for file 03.jpg**

This information can be combined to reconstruct a direct link, which will result in the file associated with the `mId` being downloaded. For example, the URL: https://mobile-api.box.com/api/1.0/download/u5es7xli4xejrh89kr6xu14tks6grjn3/2072716499 can be used to recover the JPEG image 03.jpg from the dataset. This information is not unique to the Android, and the data needed to reconstruct the URL can also be recovered from the iPhone device. Relevant artefacts can be found in the `BoxCoreDataStore.sqlite` database in the directory `/Documents/`. The privacy and legal implications associated with this practice are out of scope for this chapter.

The ultimate goal of a forensic investigator should be to recover as much evidence as possible from a cloud storage service. An analysis of the files recovered from two of the Android applications (Box and Syncplicity) has revealed that different files were recovered from different versions of these cloud storage applications. Forensic toolkits recovered fifteen files from version 1.6.7 of the Box application and only six from version 2.0.2, while five files were recovered from version 1.7 of Syncplicity application and fifteen from version 2.1.1. These results suggest that there is an opportunity to recover a more complete dataset from the cloud service if multiple devices are examined as part of an investigation. The results from the experiment propose the idea that an investigator who analyses multiple devices with different versions of an application could recover a more complete dataset than that from just a single device. The proportion of artefacts which can be recovered from two or more devices are calculated as $|m_1 \cup m_2 \cup m_3|$, where $m_n$, are the devices which are being analysed as part of a forensic investigation of cloud storage services. Preliminary data demonstrated in Table 7 – Total Files Recovered from Multiple Devices supports the idea that multiple devices can produce a more complete dataset for a forensic investigator. In three out of the four applications examined, a bigger dataset was recovered by combining the number of files recovered from each device to create a more complete dataset.

Finally, the results from the experiment can also be used to support the hypotheses proposed in Section One. $H_1$, the smartphone devices in this experiment contain a partial view of the data held in the cloud storage service. This statement continues to hold when the device is powered down. Therefore, a smartphone device potentially presents a forensic investigator with a proxy view of the evidence held in the cloud storage service. In support of $H_2$, clearing the application's cache has an adverse effect on evidence collection. The data indicates partial support for $H_3$ in that different files are recovered from the same cloud application on different mobile device platforms and operating systems for some cloud applications.

| Cloud Storage Service | m1 {files recovered} | m2 {files recovered} | m3 {files recovered} | \|m1 ∪ m2 ∪ m3\| {total files recovered} |
|---|---|---|---|---|
| *Dropbox* | {2,6,10,13,14, 16,17,18,20} | {2,6,10,13,14, 16,17,18,20} | {2,6,10,13,14, 17,18} | {2,6,10,13,14,16,17,18, 20} = **9** |
| *Box* | {1,2,4,5,6,8,9, 10,12,13,14,16, 17,18,20} | {5,6,8,9,10,12} | {2,6,10,14,18} | {1,2,4,5,6,8,9,10,12,13, 14,16,17,18,20}= **15** |
| *SugarSync* | {1,2,4,6,10,13, 14,16,17,18, 20} | {1,2,4,6,10,13, 14,16,17, 18,20} | {1,2,4,5,6,8,9, 10,12,13,14, 16,17,18,20} | {1,2,4,5,6,8,9,10,12,13, 14, 16,17,18,20} = **15** |
| *Syncplicity* | {2,6,10,14,17, 18,20} | {1,2,4,5,6,8,9, 10,12,13,14, 16,17,18} | {1,2,4,5,6,8,9, 10,13,14,16, 17,18,20} | {1,2,4,5,6,8,9,10,12,13, 14, 16, 17,18,20} = **15** |

**Table 7 – Total Files Recovered from Multiple Devices**

## 6. Conclusions and Future Work

The attractiveness of cloud computing is impacting where individuals and organizations store their data. The growing popularity of cloud storage services means that such environments will become an attractive proposition for cybercrime. This could result in an increase in demand for investigations of cloud storage services. However, the issue of conducting digital forensic investigations of cloud computing environments is an increasingly challenging and complex task. One of the biggest challenges facing investigators is the ability to identify and recover digital evidence from the cloud in a forensically sound manner. The remote and distributed nature of cloud computing environments means that the traditional offline approach to forensic evidence acquisition is invalidated. As a result, both industry and academia are beginning to examine different methods and techniques to investigate cloud computing environments.

This work presents the examination of end-devices such as smartphones, which have been used to access cloud storage services. The data recovered from these devices can be used by investigators as a proxy for potential evidence stored in cloud storage services. The effectiveness of this method is dependent on the operating system, specific cloud storage application implementation and usage patterns. In other words, the potential recovery of data increases if a device has been used to view the files through a cloud storage application and the user has not attempted to clear the cache of recently viewed files.

Two advantages become apparent to using this investigative approach. First, the investigator can begin the chain of custody process when the device is seized, and does not need to rely on the cloud provider to begin this process. Second, the tools and methods which have been used to recover data stored in cloud storage services are widely used by the forensic community. The recovery of metadata artefacts from the smartphone device can, in some scenarios, provide the investigator with insight into further data stored in a cloud service. The information recovered can also help justify a court order requesting assistance from the cloud storage provider to recover further files from the specific account.

Future research needs to be conducted to extend the analysis of smartphone hardware and operating systems, to increase the size and file types of the dataset and conduct research into other cloud storage services. The methodology proposed in this chapter can be extended to other smartphone devices and operating systems such as Windows Mobile and Blackberry devices. In addition, research can also be conducted to investigate other cloud storage services such as Google Drive, OneDrive and CloudMe. The dataset used in future experiments can also be extended to include additional data types as well increasing the overall number of files and files of varied sizes.

The analytical findings from this research indicated that examining multiple devices and multiple versions of cloud storage applications can result in a more complete dataset being recovered. This experiment can be extended to examine a number of mobile devices such as tablets, iPads, iPods and eBook readers. Other research questions that warrant investigating include the examination of usage patterns along with the construction of relevant timelines across multiple devices and cloud applications.

From a corporate security perspective, future work needs to examine the risk of data leakage that cloud storage applications can introduce to an organization. This research identifies the implications from a corporate policy perspective and determines if cloud applications introduce opportunities for data leakage in organizations. If so, what is the most effective way to minimize risk and maximize employee productivity? The results from this research provide the foundation for further development of security measures and policies for both cloud providers and smartphone users that mitigate the potential risk of data leakage.

## Appendix A: Metadata artefacts recovered Dropbox service

| OS | Filename | Fields |
|---|---|---|
| **Android** | db.db | **_data:** path gives the location of where the file can be recovered from the device.<br><br>**modified:** date and time file was uploaded to Dropbox service.<br><br>**is_favorite:** boolean field which indicates if file has been saved as a 'favorite', i.e. offline viewing.<br><br>**parent_path:** parent directory for the file, root directory is the default.<br><br>**last_modified:** last date and time the file was open/ modified on the device, stored as a UNIX epoch timestamp.<br><br>**display_name:** contains the name of the file as stored in the storage service.<br><br>**local_hash:** MD5 hash of file. |
| **iOS** | Dropbox.sqlite | **ZFAVORITE:** boolean field which indicates if file has been saved as a 'favorite'.<br><br>**ZSIZE:** size of the file in bytes<br><br>**ZVIEWCOUNT:** number of times file has been viewed using the device<br><br>**ZISTHUMBNAIL:** boolean field which indicates if a thumbnail exists for the file.<br><br>**ZLASTVIEWEDDATE:** date and time file was last viewed stored in MAC Absolute time<br><br>**ZPATH:** path and file name for particular file |

## Appendix B: Metadata artefacts recovered Box service

| OS | Filename | Fields |
|---|---|---|
| **Android** | json_static_model_ emailaddress_0 | **mThumbnail:** The URL of the thumbnail image of the file<br>**mFileName:** Name of the file as stored in the Box service<br>**mSha1:** SHA1 hash of the file<br>**mUpdated:** UNIX epoch timestamp which states the last time the file was updated, in this experiment it is the last time the file was last viewed on the device.<br>**mId:** Unique ID number assigned to each file<br>**mSize:** Size of the file in bytes<br>**mCreated:** UNIX epoch timestamp which states when the file was created, in this experiment this is the time when the file was uploaded and stored in the Box service.<br>**mShared:** Boolean (True/False) filed which indicates if file has been shared. |
| **iOS** | BoxCoreDataStore.sqlite | **ZBOXID:** unique ID number assigned to each file stored in the Box service account.<br>**ZSIZE:** size of the file in bytes.<br>**ZFAVORITEOBJECT:** boolean field which indicates if file has been saved as a 'favorite', i.e. offline viewing.<br>**ZUPDATED:** absolute timestamp showing when file was last updated<br>**ZLASTDOWNLOADDATE:** absolute timestamp showing when file was last downloaded to device<br>**ZCREATIONTIME:** absolute timestamp showing when file was stored in Box service<br>**ZNAME:** name of file<br>**ZSHA1:** SHA1 hash of file in Box service<br>**ZLOCALURLSTRING:** directory location for file stored on the device<br>**ZSTREAMINGURLSTRING:** URL location for file which can be accessed from Box service<br>**ZLOCALSHA1:** SHA1 hash of file on device |

## Appendix C: Metadata artefacts recovered Syncplicity service

| OS | Filename | Fields |
|---|---|---|
| **Android** | CacheDatabase.sqlite | **fileId:** unique ID number assigned to each file stored in the service.<br>**name:** name of file.<br>**length:** size of the file in bytes.<br>**fileStatus:** boolean value which indicates if file is still stored in the service, if the value is 1 then file is still stored in service, if value is 0 then file has been deleted.<br>**thumbnailURL:** if file has a thumbnail, this is a working URL to the thumbnail stored in the service. |
| | Virtual_File_System.db | **File_ID:** unique ID number assigned to each file stored in the service.<br>**File_Name:** name of file.<br>**Is_Favorite:** boolean field which indicates if file has been saved as a 'favorite', i.e. offline viewing.<br>**Server_Length:** size of file stored in service, presented in bytes.<br>**Local_Length:** size of file stored in device, presented in bytes.<br>**Is_Deleted:** boolean field which indicates if file has been deleted.<br>**Thumbnail_URL:** if file has a thumbnail, this is a working URL to the thumbnail stored in the service. |
| **iOS** | syncplicity.sqlite | **ZLENGTH:** size of file in bytes<br>**ZFILEID:** unique ID number assigned to each file stored in the service.<br>**ZDELETED:** boolean field which indicates if file has been deleted.<br>**ZFILENAME:** name of file.<br>**ZEXT:** file type.<br>**ZTHUMBNAILURL:** if file has a thumbnail, this is a working URL to the thumbnail stored in the service |